\documentclass[%
 aip,
 amsmath,amssymb,
 reprint,%
]{revtex4-1}

\usepackage{graphicx}
\usepackage{dcolumn}
\usepackage{bm}

\usepackage[utf8]{inputenc}
\usepackage[T1]{fontenc}
\usepackage{mathptmx}

\usepackage{color}
\usepackage{soul}

\begin{document}

\preprint{AIP/123-QED}

\title[Phase Transition in Iterated Quantum Protocols for Noisy Inputs]{Phase Transition in Iterated Quantum Protocols for Noisy Inputs}

\author{Martin Malachov}
\author{Igor Jex}%
\affiliation{ 
Faculty of Nuclear Sciences and Physical Engineering, Czech Technical University in Prague \\
B\v rehov\'a 7, 115 19 Praha 1 - Star\'e M\v esto, Czech Republic
}%

\author{Orsolya K\'alm\'an}
\author{Tam\'as Kiss}
\affiliation{%
 Institute for Solid State Physics and Optics, Wigner Research Centre for Physics\\
Hungarian Academy of Sciences, Konkoly-Thege \'ut, H-1121 Budapest, Hungary
}%

\date{\today}

\begin{abstract}
Quantum information processing exploits all the features quantum mechanics offers. Among them there is the possibility to induce nonlinear maps on a quantum system by involving two or more identical copies of the given system in the same state. Such maps play a central role in distillation protocols used for quantum key distribution. We determine that such protocols may exhibit sensitive, quasi-chaotic evolution not only for pure initial states but also for mixed states, i.e. the complex dynamical behavior is  not destroyed by small initial uncertainty. 
We show that the appearance of sensitive, complex dynamics associated with a fractal structure in the parameter space of the system has the character of a phase transition. The purity of the initial state plays the role of the control parameter and the dimension of the fractal structure is independent of the purity value after passing the phase transition point.
The critical purity coincides with the purity of a repelling fixed point of the dynamics and we show that all the pre-images of states from the close neighborhood of pure chaotic initial states have purity larger than this. Initial states from this set can be considered as quasi-chaotic.   
\end{abstract}

\maketitle

\begin{quotation}
Quantum informational protocols involve coherent evolution, measurement and postselection of qubits. A typical example is entanglement distillation. The resulting conditional dynamics is nonlinear, in contrast to the usual evolution of both closed and open quantum systems. Already the simplest types of such protocols may result in rich, complex chaotic dynamics when applied iteratively. An important property of these iterated dynamical systems is that initially pure quantum states remain pure throughout the evolution. For single-qubit systems, there is a one to one correspondence of the pure-state quantum dynamics to the iterated dynamics of quadratic rational maps with one complex variable. The initial states not converging to stable cycles form a closed set on the Bloch sphere, corresponding to the Julia set of the quadratic rational map. This set, describing the chaotic regime of the dynamics, may have a fractal structure. We investigate here the problem of starting the dynamics of a single-qubit protocol from an initially noisy quantum state, described by a density matrix. The initial noise either diminishes during the evolution, resulting in the purification of the quantum state, or it gets amplified by the process, resulting in the completely mixed state. 
We carry out a systematic analysis and present strong evidence that the fractal structure of the border points between different convergence regions remains a fractal for noisy initial states, up to a certain critical purity value. The dimension of the fractal, estimated by box counting, remains constant and drops suddenly at the critical purity. This is analogous to a phase transition. An alternative way to examine the nature of this transition point is to follow the pre-images of points from the neighborhood of the pure-state fractal. 
We find that these pre-images always fall on borders between convergence regions. They never have purity smaller than the critical purity which coincides with the purity of an unstable fixed point 
to which branches of the backwards iterated fractal points 
converge, verifying the transition point. 
\end{quotation}

\section{\label{sec:level1}Introduction
}

The effect of noise is a threat to all information processing applications of quantum mechanics. Too much external noise or, with other words, randomness destroys quantum coherence and harms the successful operation of a quantum machine. 
Quantum computers realize quantum information protocols and hence rely on the repetition (iteration) of identical operations. Essential features of such repeated iterations can be uncovered by examining simple setups.
In order to study noise in quantum computation on an elementary model, we assume that all essential ingredients of quantum information processing are available: initial state preparation, coherent manipulation and gaining information by measurement at the output.
 Any possible quantum computer capable of coherent manipulation of qubits is known to be reducible to an arrangement of identical two-qubit gates (e.g. a controlled NOT gate)  and single-qubit gates. Perhaps the simplest arrangement of these elements is a CNOT gate followed by one single-qubit gate with an adjustable parameter and a von Neumann measurement on one of the output qubits.  We will apply this arrangement on two qubits from an ensemble of identically prepared qubits and measure one of the output qubits \cite{Gisin} and then simply repeat the same procedure. Studying such a protocol is not a mere curiosity or mathematical exercise. The protocols under discussion are known as distillation protocols and are used for several purposes. They can be applied to increase the purity of entangled states in such a way as to increase their degree of entanglement which is a resource in quantum information processing. Another practical application of such protocols is to drive apart almost identical quantum states making them much better identifiable in a quantum mechanical sense which is a critical step in information retrieval and hence in quantum information processing. As a result of the inherently nonlinear nature of these protocols, chaos may appear. The conditions leading to chaotic evolution is an important aspect worth studying as such a regime may seriously impair the practical applicability of a protocol for information processing.

The peculiar features of these protocols lie in the systematic, repeated application of the same set of operations on pairs of states that have been transformed in the same way in the previous steps. This  iterative application of the same quantum transformation leads to surprisingly rich dynamics, where not only complex chaos appears \cite{Kiss2006}, or the convergence regions to stable cycles of the dynamics in the space of initial states may become fractal bordered \cite{Kiss2008}, but also the inherent true sensitivity to initial conditions can be employed for useful purposes e.g. for quantum state discrimination \cite{Paris,Torres,Kalman}.
Multi-qubit schemes based on similar operations can be employed for entanglement distillation \cite{Bennett97,Deutsch98,Alber,Kiss2011}. None of these features would be possible by undisturbed unitary dynamics \cite{Ballentine,Weinstein} or even in an open quantum system governed by a completely positive map, as they act linearly on superpositions of initial states. It is the postselection based on measurement results after each iterational step which gives rise to nonlinear behavior \cite{Terno1999,Totharxiv,Filippov2017} and leads to the above sensitive phenomena. But how sensitive are these effects to external noise? One may na\"\i vely expect that for noisy initial states the fine details of the fractal structures characterizing the border between convergence regions in the space of initial states would be washed out \cite{Guan2013,Kalmanarxiv}. 

For pure state inputs, measurement induced iterative dynamics of qubits can be effectively described by iterated maps with one complex variable, having the form of a quadratic polynomial or a quadratic rational function. It has been proven that by replacing the CNOT gate by another appropriate unitary gate, we can implement any quadratic rational function (including quadratic polynomials). Moreover, any rational function of arbitrary degree can be realized by higher dimensional unitary gates with more input qubits \cite{Andras}. Thus it is not a surprise that the operation of the above scheme for pure initial states exhibits high resemblance to the well-known and extensively studied logistic map of chaos theory, being itself an iterated quadratic polynomial with one real variable \cite{Log1,Log2,Milnor}.  Among the basic features of such maps we can list the appearance of period doubling, the appearance of limit cycles of high degree and the appearance of fractal structures specifying the boundary between different asymptotic stationary regimes of the system.
For noisy (mixed state) inputs, however, the dynamics does not reduce to a map with one complex variable, therefore we can expect the occurrence of novel phenomena not included in previous studies which have been limited primarily to initial pure states which assumption considerably simplified the analysis. 

In this paper we prove that the characteristic fractal structure of the dynamics survives for initially noisy (mixed) states. We numerically estimate its fractal dimension and prove that it is independent of the initial purity above a specific threshold which is defined by a critical initial purity of the system. Below the critical purity, we find that the fractal dimension drops. This sudden change has the property of a phase transition \cite{Sole1996} where the phase of the component is characterized by the presence of a fractal structure.  Additionally, we study which noisy initial states have a close connection with the pure-state fractal and can thus be termed ''quasi-chaotic''.

\section{Nonlinear protocol}
We consider the iterations of the elementary nonlinear scheme by Bechmann-Pasquinucci, Huttner and Gisin \cite{Gisin} with an additional single qubit unitary Hadamard gate   
\begin{equation}\label{protocol}
\rho\rightarrow U_{H} \frac{\rho\odot\rho}{\text{Tr}(\rho\odot\rho)} U_{H}^{\dagger} \, ; \quad U_{H}=\frac{1}{\sqrt{2}} \begin{pmatrix} 1 & 1 \\ 1 & -1 \end{pmatrix},
\end{equation}
where the $\odot$ symbol stands for element wise (Hadamard) product in the computational basis.  The density matrix can be parameterized with its Bloch-sphere coordinates $u,v,w\in \mathbb{R}$ as
\begin{equation}
\label{1qstate}
\rho=\frac{1}{2}\begin{pmatrix}1+w & u-iv  \\ u+iv & 1-w \end{pmatrix} 
\end{equation}
and the purity of the state is given by $P=\text{Tr}(\rho^2)=(1+u^2+v^2+w^2)/2\leq 1$, which leads to the condition $u^2+v^2+w^2\leq 1$.
After one step of the protocol the parameters describing the state are transformed by the nonlinear map $\mathcal{M}:\mathbb{R}^{3}\to\mathbb{R}^{3}$ into
\begin{equation}\label{evolution}
u'=\frac{2w}{1+w^2}, \quad v'=\frac{-2uv}{1+w^2}, \quad w'=\frac{u^2-v^2}{1+w^2}.
\end{equation}
The analysis of $\mathcal{M}$ reveals that the $\left(u,0,w\right)$ plane is an invariant subset of the map, where all the four quadrants are mapped to the first quadrant within at most two steps in a counterclockwise manner (c.f. Fig~\ref{Fig1}a). Three of the fixed points of ${\cal M}$ ($C_{0}$, $C_{1}$, and $C_{2}$) and both of its length-2 cycles ($C_{3}$ and $C_{4}$) lie on this plane (see Table~\ref{fixed_cyc}).  $C_{0}$ and $C_{3}$ are attractive, while $C_{1}$ is repelling in every direction, this can be shown by examining the second iteration for a small neighborhood. Cycles $C_{2}$ and $C_{4}$ exhibit both attractive and repelling behavior depending on the direction. Furthermore, $\mathcal{M}$ has two other fixed points, $C_{5}$ and $C_{6}$ (see Table~\ref{fixed_cyc}), which do not lie on the  $\left(u,0,w\right)$ plane. According to our numerical analysis, the map $\cal M$ does not have any globally attractive cycles beyond $C_{0}$ and $C_{3}$. We note that longer cycles of $\cal{M}$ can be found by determining the roots of multivariate polynomials of high order, which, even  numerically, is challenging. Our calculations have not found longer cycles of $\cal{M}$ that would not coincide with those of the complex rational function which describes the pure-state dynamics (see below), and are not contained in the Julia set of the latter.

\begin{table}[h!]
\centering
\begin{tabular}{|c|c|c|}
 \hline
 \multicolumn{2}{|c|}{Fixed points \& cycles of $\mathcal{M}$} & \multicolumn{1}{c|}{$P$} \\
 \hline
 $C_{0}$ & $\left(0,0,0\right)$ & $0.5$\\
 \hline
 $C_{1}$ & $\left(0.639,0,0.361\right)$ & $0.769$\\
 \hline
 $C_{2}$ & $\left(0.839,0,0.544\right)$ & $1$ \\
 \hline
 $C_{3}$ & $\left(1,0,0\right) \leftrightarrow \left(0,0,1\right)$ & $1$ \\
\hline
 $C_{4}$ & $\left(0.544,0,0\right) \leftrightarrow \left(0, 0, 0.296\right)$ & $0.648 \leftrightarrow 0.544$ \\
\hline
 $C_{5}$ & $\left(-0.544, 0.786, -0.296\right)$ & $1$ \\
\hline
 $C_{6}$ & $\left(-0.544, -0.786,-0.296\right)$ & $1$ \\
\hline
\end{tabular}
\caption{Fixed points and cycles of length two of the map $\mathcal M$ and the corresponding values of the purity. Note that values are rounded to three decimal places.}
\label{fixed_cyc}
\end{table}

\section{Pure-state dynamics}
In the case of $P\!\!=\!\!1$ the evolution by the map $\cal M$ does not change the purity, the states remain pure. 
For pure states one can introduce a complex variable $z$ with $\left|z\right|^2=(1-w)/(1+w)$, and $\arg(z)=\arctan(v/u)$ to parameterize quantum states as $\left|\psi\right>=\left(\left|0\right>+z\left|1\right>\right)/\sqrt{\smash[b]{1+\left|z\right|^{2}}}$. The iterative dynamics can then be described by the complex quadratic rational function \cite{Milnor,MilnorGD,Kiss2011,Andras,Torres} $f_{\cal M}(z)=(1-z^2)/(1+z^2)$. The fixed points $C_{2}$, $C_{5}$ and $C_{6}$ are part of its Julia set, while $C_{3}$ is its single attractive fixed cycle (c.f. Fig. 1b) corresponding to the states $\big|\psi_{3}^{(1)}\big>=\left(\left|0\right>+\left|1\right>\right)/\sqrt{2}\leftrightarrow 
\big|\psi_{3}^{(2)}\big>=\left|0\right>$. Single-qubit pure states which converge to this attractive cycle form the so-called Fatou set of the map, all other pure states belong to the Julia set containing also the repelling cycles, which are everywhere dense in it. Pure initial states that belong to the Julia set remain in the Julia set under the iteration of $\cal M$. The Julia set is formed at the border between the different Fatou-components -- in this case at the border between sets of points which converge to $C_3^{(1)}$ or to $C_3^{(2)}$ after an even number of iterations (Fig.~\ref{Fig1}b). The Julia set is a fractal, its Hausdorff dimension can be calculated e.g. by the periodic point algorithm it is estimated\cite{Andras_private_comm} to be $1.566$. Since the Julia set is nonvacuous \cite{Milnor} the protocol for pure states has chaotic dynamical regimes.


\section{Phase transition for noisy inputs} 
\begin{figure}
\includegraphics[width=0.235 \textwidth]{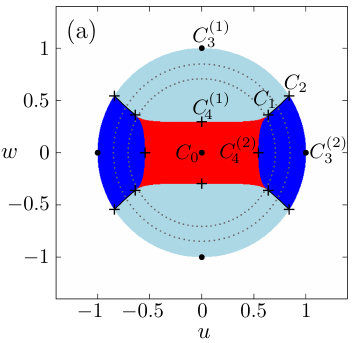}
\includegraphics[width=0.235 \textwidth]{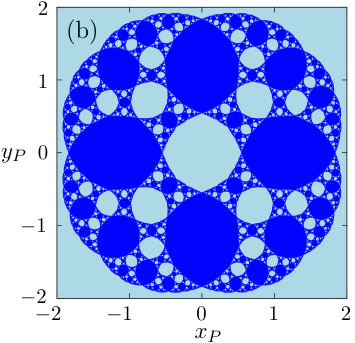} \\
\includegraphics[width=0.235 \textwidth]{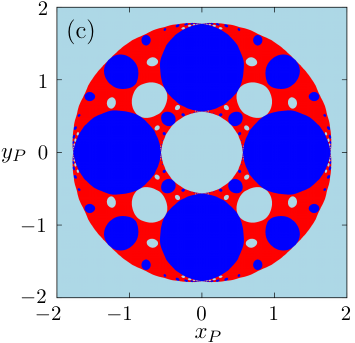}
\includegraphics[width=0.235 \textwidth]{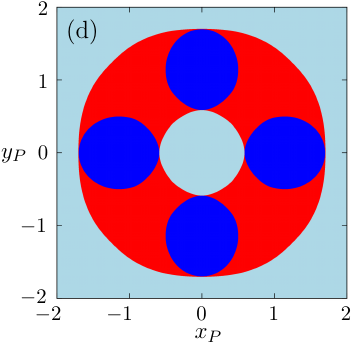}
\caption{(color online) (a) The regions of convergence of the map $\cal M$ in the $(u,0,w)$ plane of the Bloch sphere. Initial states represented by red color converge to the attractive cycle $C_{0}$, light blue and dark blue regions represent states which converge after an even number of iterations to $C_{3}^{(1)}$ and $C_{3}^{(2)}$, respectively. Attractive fixed cycles and their pre-images are shown by black dots, the other fixed cycles (and their pre-images) are denoted by black crosses. The black curves contain points which converge unstably from $C_{1}$ to $C_{2}$ along the curve. Gray dotted circles correspond to initial purity values considered in (c) and (d). (b)-(d) stereographic projections  of spherical surfaces corresponding to initial purity values $P=1$, $0.87$ and $0.75$, respectively. Coloring is the same as in (a). The stereographic projections are taken from the south pole of the spherical surface onto the equatorial planes and the axes are given as $x_{P}=u/(\sqrt{2P-1}+w)$ and $y_{P}=v/(\sqrt{2P-1}+w)$.  
}
\label{Fig1}
\end{figure}

Considering noisy (i.e., mixed) initial states, there are three characteristically different behaviors: an initial state can converge to either one of the two globally attractive cycles $C_{0}$ and $C_{3}$ or not converge stably to any of these. 
Since $C_{0}$ corresponds to the completely mixed state $\rho_{0}=(\left|0\right>\left<0\right|+\left|1\right>\left<1\right|)/2$, while $C_{3}$ corresponds to the pure-state cycle $\big|\psi_{3}^{(1)}\big>\leftrightarrow\big|\psi_{3}^{(2)}\big>$, the question naturally arises whether a general state with purity $1/2 < P < 1$ will become pure or completely mixed by the iteration of the protocol (see Fig.~\ref{Fig1}(a)). It turns out that initial states with the same purity and close to each other may converge to the same or to different attractive cycles according to some complicated pattern. In Fig.~\ref{Fig1}(c)-(d) we present stereographic projections of surfaces corresponding to initial states with a given $P$ (with the respective value represented by the dotted circles on Fig.~\ref{Fig1}(a)), colored according to which cycle the states converge to.

\begin{figure}
\includegraphics[width=0.47\textwidth]{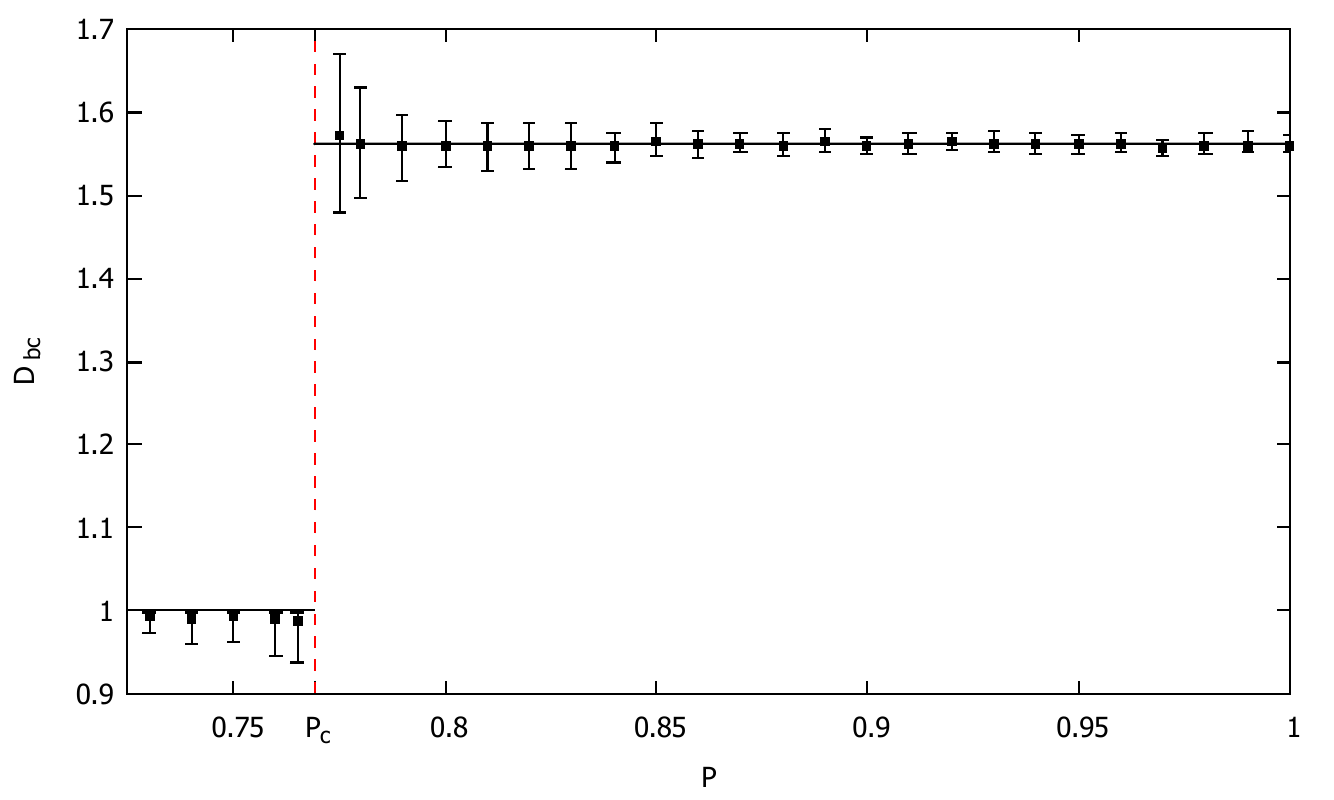}
\caption{
The box-counting fractal dimension $D_{\mathrm{bc}}$ of the border between the different convergence regions as a function of the initial purity  \cite{BCM}. The error bars indicate the range of numerically calculated values $D_{\mathrm{bc}}$ obtained by rotating the grid of the box-counting method with respect to the fractal. The black squares correspond to the mean value of the 37 data points for each initial purity. For initial purities above $0.78$,  $D_{\mathrm{bc}}$ is fluctuating around the average value $1.561$ (solid line). For $P<0.76$, where the structure is no longer a fractal, $D_{\mathrm{bc}}$ equals $1$ within the statistical error of the numerical method.  The numerical precision of our implementation of the box-counting method does not provide reliable results for $0.765<P<0.775$.} 
\label{dimension}
\end{figure}

As the purity of initial states is decreased below $1$, in addition to the pure-state cycle $\big|\psi_{3}^{(1)}\big>\leftrightarrow\big|\psi_{3}^{(2)}\big>$, one can find states which converge to the maximally mixed state $\rho_{0}$. 
Thus the presence of noise in the initial quantum state can lead to a different output of the protocol than one would get for an ideal pure-state input. However, the border between states converging to the different attractive cycles preserves its fractalness. Fig.~\ref{Fig1}(c)  shows the case when the fractal is still visible with the naked eye at the same magnification and resolution as the pure-state fractal. For lower values of the purity, the fractal further shrinks and can only be seen under magnification, our numerical simulations indicate that it is present only above a certain critical initial purity which we found to be $P_{\mathrm{c}}= 0.769$ ($P_{\mathrm{c}}$ is equal to the purity $P_{1}$ of the fixed point $C_{1}$ up to this numerical precision). Furthermore, we have calculated the dimension of the fractals using the box counting method \cite{BCM} for different $P$ values, as shown in Fig.~\ref{dimension}. The fractal dimension approximately remains the same value as for $P=1$ until the purity is decreased to $P_{\mathrm{c}}$, where it suddenly drops to the value $1$, meaning the convergence regions become regular (i.e., non-fractal) below this value. This effect resembles a phase transition where the purity plays the role of the control parameter and the sudden appearance of a fractal at the critical purity with a jump in the fractal dimension marks the transition point \cite{Sole1996}.

\section{Quasi-chaotic noisy initial states} 
In the case of pure inputs corresponding to the Julia set (points lying on the border of light and dark blue convergence regions in Fig.~\ref{Fig1}b), iterations are numerically unstable and lead out of the set due to numerical errors. However, under the inverse map the Julia set is attractive and iterations are numerically stable i.e., starting from a point belonging to the Julia set, the numerically determined pre-images will also result points, which belong to the Julia set. In the case of the more general map $\cal{M}$ unstable behavior is seen for all points which lie on some border between the different convergence regions. Points on the invariant plane which lie on the border of light and dark blue colors in Fig.\ref{Fig1}(a) unstably converge to $C_{2}$, while points lying on the border of red and blue colors unstably converge to the length-2 cycle $C_{4}$. In fact, this unstable forward convergence can be proven by applying the inverse map (i.e., backwards iteration) on these points, as the reversed dynamics results in a stable reversed convergence. With backwards iteration we have two pre-images in each step (${\mathcal M}^{-1}_{+}$ and ${\mathcal M}^{-1}_{-}$) thus we can choose which branch we consider\cite{Kalmanarxiv}. There are two special branches: $\left({\mathcal M}^{-1}_{+}\right)^{\circ \, n}$ and $\left({\mathcal M}^{-1}_{-}\right)^{\circ \, n}$. Under the special branch $\left({\mathcal M}^{-1}_{+}\right)^{\!\circ\, n}$ points from the vicinity of $C_{2}$ (which is a Julia-set point on the $P=1$ surface) converge to $C_{1}$, and the other branches produce the pre-images of these points in the other quadrants (see the black curves which were numerically obtained by this method). The unstable forward convergence along the curve starting from the fixed point $C_{1}$ towards the Julia-set-member fixed point $C_{2}$ is an interesting feature of the map $\cal{M}$ and raises the question, whether there are similar points  with $P<1$ lying outside the invariant plane, which also converge unstably to some point of the Julia set, i.e., the fractal characterizing the pure-state dynamics.

\begin{figure}
\includegraphics[width=0.48\textwidth]{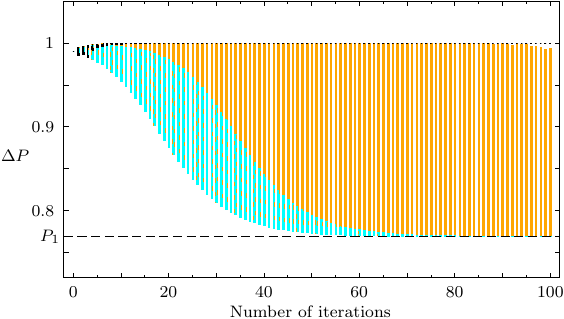}
\caption{The purity range spanned by states obtained by backwards iteration of the map ${\cal M}$.  $1024$ points were generated by taking all pre-images (up to 10 iterations) of a single point from the Julia set, resulting in quasi-randomly distributed points within the Julia set. Their radial coordinate was then reduced by applying the transformation $u\rightarrow 0.99\,u$, $v\rightarrow 0.99\,v$, $w\rightarrow 0.99\,w$, so that their purity corresponded to $P=0.99$. Black and cyan colors refer to iterations of these initial points with the two special branches  $\left({\mathcal M}^{-1}_{-}\right)^{\!\circ\, n}$ and  $\left({\mathcal M}^{-1}_{+}\right)^{\!\circ\, n}$, respectively. Orange corresponds to iterations with various randomly chosen branches. Note that the plots are superposed in the order orange$\rightarrow$cyan$\rightarrow$black. }
\label{deltaP}
\end{figure}

Let us assume that there exist some points with purity $P\!<\! 1$ that tend to the Julia set in a similarly unstable manner under forward iteration. If they exist, we can find such points numerically by applying backwards iteration to initial states close to the Julia set\cite{Kalmanarxiv}, but with purity $P\!< \!1$ . For this purpose, we have iterated randomly chosen branches of the inverse map starting from randomly chosen mixed initial states from the vicinity of the Julia set (see Fig.~\ref{deltaP}). We found that none of the resulting points have a purity smaller than $P_{1}$ (our numerical results approach $P_{1}$ from above up to a numerical deviation of $8.6\times 10^{-6}$). The special branch $\left({\mathcal M}^{-1}_{-}\right)^{\!\circ\, n}$ does not decrease the purity asymptotically, as it maps points towards its two fixed points $C_{5}$ and $C_{6}$, both corresponding to pure states. The other special branch $\left({\mathcal M}^{-1}_{+}\right)^{\!\circ\, n}$ on the other hand, tends to $P_{1}$ for any such initial state (this property is due to the fact that this branch has $C_{1}$ as a fixed point). Branches with a combination of the two types of pre-images have not been found to show any other tendency. The fact that the points from the vicinity of the fractal of the pure case thus do not have any pre-images with purity smaller than $P_{1}$, together with our results regarding the drop of the fractal dimension, and the disappearance of the fractal-like structures at approximately $P_{1}$ indicate that the critical purity $P_{\mathrm{c}}$ coincides with $P_{1}$.

\begin{figure}
\includegraphics[width=0.35 \textwidth]{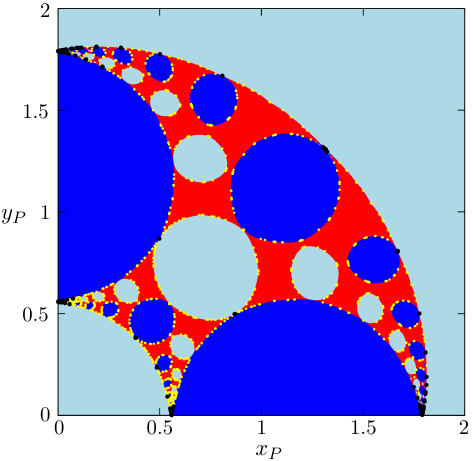}
\caption{The first quadrant of the stereographic projection of the $P=0.9$ surface colored according to the converge of initial states. Blue and red colors refer to convergence to the stable cycles, black (yellow) dots correspond to the pre-images of points initially in the vicinity of the Julia set (the red-blue border points on the invariant plane).}
\label{Fig4}
\end{figure}

Using the inverse map ${\mathcal M}^{-1}$, one can look for pre-images of border points on surfaces of different purity. 
We chose two sets of initial points which lie outside the invariant plane: (i) points in the vicinity ($P<1$) of randomly chosen points of the Julia set and (ii) points from the vicinity of the red-blue border on the $(u,0,w)$ plane. We found a similar behavior as on the invariant plane. Namely, with the different branches of the backwards iterated map, the border points of the two blue regions remained on such type of borders, and the red-blue border points also stayed on the border between the red and blue convergence regions. This is illustrated in Fig.~\ref{Fig4}. 
   
Reversing the results of the backwards iterational method reveals that the border points of the light and dark blue regions converge (unstably) to the points of the Julia set. Since the points of the Julia set are related to the chaotic dynamical behavior of our protocol, these mixed initial states may be considered quasi-chaotic. Let us note that the states which correspond to the red-blue border points cannot be directly connected to the pure chaotic states with this method. 

\section{Conclusions}
We studied the iterated dynamics of a particular measurement-induced nonlinear quantum protocol in the presence of initial noise.  We found a sudden transition from fractal border to non-fractal border between convergence regions in the dynamics as a function of the purity of the initial states. The transition takes place at a single critical initial purity which resembles a phase transition, where the purity plays the role of the control parameter. 

We studied other properties than those usually analyzed for quantum chaos \cite{Haake}. Due to their inherent complex character, our results underline much more the high degree of complexity which iterated rational maps can exhibit. 
To look for similar phase transitions in protocols for multiqubit systems e.g. related to entanglement distillation remains a challenge. When iterated functions of more complex variables are involved, it is a formidable task to derive analytic results as essential mathematical theorems are almost absent from the literature \cite{Fornaess}. One can hope, however, that special cases may be analysed by methods similar to the one discussed in this paper. Indeed, according to our unpublished results, there is a special two-qubit protocol based on LOCC operations and capable of entanglement distillation, for which a restricted set of mixed initial states can be characterized by three real variables, which evolve according to a map similar to the one presented here.
More general situations may cover interesting physical cases for more complex gates or higher dimensional inputs.

\begin{acknowledgments}

M.M. appreciates the financial support from the Czech Technical University in Prague under Grant No. SGS16/241/OHK4/3T/14. I. J. received support from the Czech Grant Agency under grant No. GA \v CR 16-09824S and from MSMT RVO 68407700. This publication was funded by the project "Centre for Advanced Applied Sciences", Registry No. CZ.$02.1.01/0.0/0.0/16\_019/0000778$, supported by the Operational Programme Research, Development and Education, co-financed by the European Structural and Investment Funds and the state budget of the Czech Republic. T. K. and O. K. are grateful for the support of the National Research, Development and Innovation Office of Hungary (Project Nos. K115624, K124351, PD120975, 2017-1.2.1-NKP-2017-00001). O. K. acknowledges support from the J\'{a}nos Bolyai Research Scholarship of the Hungarian Academy of Sciences.

\end{acknowledgments}

\bibliography{aipsamp}

\end{document}